\documentclass[twocolumn,showpacs,preprintnumbers,amsmath,amssymb,pra,floatfix]{revtex4-1}
\usepackage{graphicx}
\usepackage{wrapfig}
\providecommand{\e}[1]{\ensuremath{\times 10^{#1}}}

\usepackage[usenames,dvipsnames]{color}

\begin{document}
\title{Analysis of spatial emission structures  in vertical-cavity surface-emitting lasers with feedback of a volume Bragg grating}

\author{Y. Noblet}
\author{T. Ackemann}
\email[]{thorsten.ackemann@strath.ac.uk} \affiliation{SUPA,
Department of Physics, University of Strathclyde, 107 Rottenrow,
Glasgow G4 0NG, Scotland, UK}


\begin{abstract}

We investigate the spatial and spectral properties of broad-area vertical-cavity surface-emitting lasers (VCSEL) with frequency-selective feedback by a volume Bragg grating (VBG). We demonstrate wavelength locking similar to the case of edge-emitters but the spatial mode selection is different from the latter. On-axis spatial solitons obtained at threshold give way to off-axis extended lasing states beyond threshold. The investigations focus on a self-imaging external cavity. It is analyzed how deviations from the self-imaging condition affect the pattern formation and a certain robustness of the phenomena is demonstrated.

\end{abstract}


\maketitle

\section{Introduction}

Volume Bragg gratings (VBG) are compact, narrow-band frequency
filters which prove to be of increasing use in photonics. One
particular application is the wavelength control of edge-emitting
laser diodes (EEL), where they can stabilize the emission wavelength
very effectively against the red-shift connected to an increase of ambient
temperature or to the Ohmic heating  due to increasing current \cite{volodin04,maiwald08,kroeger09}. In addition, the
spectral and spatial brightness of broad-area EEL can be reduced
significantly by the feedback from a VBG \cite{volodin04,maiwald08}.
Commercial versions are referred to as `wavelength-locker' or
`power-locker'.

Frequency-selective feedback was proposed and demonstrated to
provide some control of transverse modes also for vertical-cavity
surface-emitting lasers (VCSEL), both for medium-sized devices
emitting Gaussian modes \cite{marino03,chembo09} as well as for
broad-area devices emitting Fourier modes
\cite{schulz-ruhtenberg09}. These investigations used diffraction
gratings and to our knowledge the only investigations using VBG were
performed with a focus on the close-to-threshold region where
bistable spatial solitons are formed \cite{radwell09,ackemann09a}.
In this work, we are reporting experiments on how the solitons
formed at threshold give way to spatially extended spatial
structures and analyze their properties quantitatively. We will
discuss similarities and differences to the edge-emitting case and
to what extend VBG are helpful to control spatial modes in a VCSEL.
We are focusing on a specific setup of the external cavity close to
a self-imaging situation and analyze how deviations from the
self-imaging condition are influencing the pattern formation.

\section{Experimental Setup}

\begin{figure}[t!]
     \centerline{
     \includegraphics[scale=0.25,angle=0]{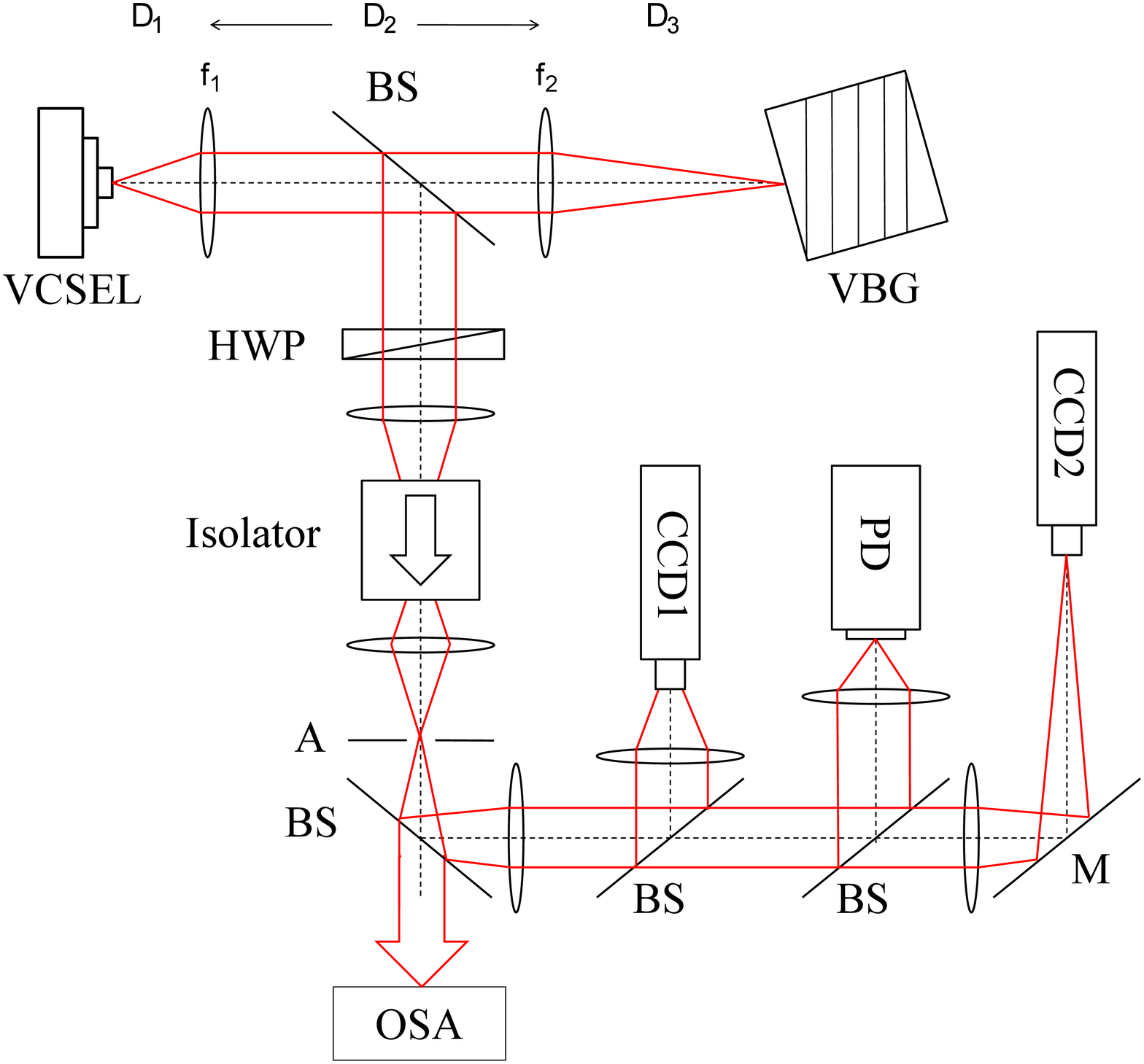}
     }
     \caption{Experimental setup. VCSEL: Vertical-cavity surface-emitting laser, BS: Beamsplitter, VBG: Volume Bragg grating, HWP: Half wave plate, A: Aperture, M: Mirror, PD: Photodiode, CCD1: CCD camera in near field image plane of VCSEL, CCD2: CCD camera in far field image plane of VCSEL, OSA: Optical spectrum analyzer.}
    \label{fig:setup}
     \end{figure}

A schematic diagram of the experimental setup is illustrated in
Fig.~\ref{fig:setup}. The VCSEL used for this experiment is
fabricated by Ulm Photonics and similar to the ones described in
more detail in
\cite{grabherr98,grabherr99,radwell09,schulz-ruhtenberg09}. It is a
large aperture device, allowing for the formation of many transverse cavity
modes of fairly high order, with a $200~{\rm \mu m}$ diameter
circular oxide aperture providing optical and current guiding. The emission takes place through the n-doped Bragg reflector and through a transparent substrate (so-called bottom-emitter, \cite{grabherr98,grabherr99}).
The laser has an emission wavelength around $975~{\rm nm}$ at room
temperature. The VCSEL is tuned in temperature up to $70~{^\circ\rm
C}$ so the emission wavelength approaches the reflection peak of the
volume Bragg grating (VBG). The VBG has a reflection peak at
$\lambda_{\rm g}=981.1~{\rm nm}$ with a reflection bandwidth of $0.2~{\rm nm}$
full-width half-maximum (FWHM). At such a high temperature the free
running laser nearly has an infinite threshold and lasing only
occurs because of the feedback from the VBG.

The VCSEL is coupled to the VBG via a self-imaging external cavity.
Every point of the VCSEL is imaged at the same spatial position
after each round trip therefore maintaining the high Fresnel number
of the VCSEL cavity.
The VCSEL is collimated by $f_1=8~{\rm mm}$ focal length
plano-convex aspheric lens. The second lens is a $f_2=50~{\rm mm}$ focal
length plano-convex lens and is used to focus the light onto the VBG.
This telescope setup gives a $6.25:1$ magnification factor onto the
VBG. This cavity has a round trip frequency of $1.23~{\rm GHz}$
which corresponds to a round trip time of $0.81~{\rm ns}$. The light
is coupled out of the cavity using a glass plate (beam splitter with
a front uncoated facet and a back anti reflection coated facet). The
reflection is relying on Fresnel reflection and therefore is
polarization dependent. The reflectivity is on the order of $10~{\rm
\%}$ for s-polarized light and $1~{\rm \%}$ for p-polarized light.

An optical isolator is used to prevent reflection from the detection
to pass into the external cavity. There are two
charge-coupled-device (CCD) cameras used for detection, one is used
to produce images of the VCSEL gain region (near field) and the
other camera produces images of the Fourier plane of the gain region
(far field). The emission spectrum is recorded with an optical
spectrum analyzer (OSA). There is also a photodiode which measures
the laser power.

\begin{figure}[thb]
\includegraphics[width=80mm,clip=]{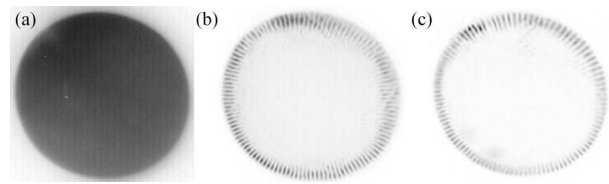}
\caption{ Near field intensity distribution of the VCSEL showing a)
spontaneous emission at $I=200~{\rm mA}$  (at a temperature of
$70~{^\circ\rm C}$) without feedback (this, and all other images in
this article, depict intensity in a linear gray scale with black
denoting high intensity; b) VCSEL without feedback, $I=490~{\rm
mA}$ at $T=16~{^\circ\rm C}$; c) VCSEL with feedback from a plane mirror, $I=185~{\rm mA}$ at $T=70~{^\circ\rm C}$.
Note that the slight non-homogenous behavior of the spontaneous
emission (decrease of intensity from the upper left to the lower
right) is due to the detection setup.} \label{fig:VCSEL}
\end{figure}

The spontaneous emission is rather homogeneous below threshold
(Fig.~\ref{fig:VCSEL}a) but as the current increases the intensity
becomes higher at the boundaries due to current crowding at the
oxide aperture \cite{grabherr99}. In spite of the large aperture the
device is still capable of lasing at low enough temperatures. The
lower the temperature the lower the current required to achieve
lasing. Above threshold the laser starts to lase on a kind of
whispering gallery mode (e.g.\ \cite{schulz-ruhtenberg09a}) around
the boundaries of the VCSEL (Fig.~\ref{fig:VCSEL}b).  The gain is
the highest due to the current crowding previously observed
therefore leading to a lower threshold of this mode.

\section{Experimental results}

\subsection{Alignment of self-imaging cavity}

As indicated, the cavity consists of two lenses forming an
astronomical telescope. Thus the magnification factor M of this
telescope is determined by the ratio of the two focal lengths. The
position of the first lens, $D_1$,  is adjusted to provide the best
collimation of the VCSEL output. The distance between the two
intra-cavity lenses, $D_2$ should be ${\rm f_1~+f_2}$ for an afocal
telescope, where $\rm{f_1}$ and $\rm{f_2}$ are the two focal lengths
of both lenses. In reality this is difficult to adjust because the
lenses are `thick'. We started with an approximate placement given
by the nominal focal lengths but improved this as discussed later in
Sect.~\ref{sec:adjust}. (The data given in Figs.~\ref{fig:VCSEL}c and
\ref{fig:Nearfield} correspond to the optimized position.) The
position of the VBG (or the mirror) closing the cavity at $D_3$ is
determined from images like in Fig.~\ref{fig:Nearfield} obtained at
high current. When the boundaries of the aperture are sharp and well
defined this is supposedly the self-imaging position (compare the
left (and right) column to the two central ones). We remark that the
telescope is still imaging the intensity distribution for correct
$D_1$, $D_3$ but incorrect $D_2$ (see the discussion around
Eq.~\ref{eq:D2} later), but that for self-imaging of intensity and
phase profiles, $D_2$ needs to equal $f_1+f_2$.

\subsection{Feedback with a mirror}

For completeness, we study first the case of feedback with a plane
mirror, i.e.\ without frequency selection. The laser starts at a
quite low threshold of around $170~{\rm mA}$. The emission
(Fig.~\ref{fig:VCSEL}c) is characterized by a ring with fringes
perpendicular to the aperture and this is very similar to what we
observe with the free running laser at lower temperature. Obviously,
this preference for the perimeter is again a gain effect due to the
current crowding.

\begin{figure}[thb]
\begin{picture}(85,100)
\put(20,100){\bf {\rm \textbf{D$_3$}}} \put(30,100){\bf {\rm
\textbf{D$_3-4~$mm}}} \put(50.5,100){\bf {\rm \textbf{D$_3+4~$mm}}}
\put(72,100){\bf {\rm \textbf{Mirror}}} \put(0,86.5){\bf 391~mA}
\put(0,70.5){\bf 405~mA} \put(0,54.5){\bf 450~mA} \put(0,38.5){\bf
500~mA} \put(0,22.5){\bf 550~mA} \put(0,6.5){\bf 600~mA}
\put(15,0){\includegraphics[width=70mm,clip=]{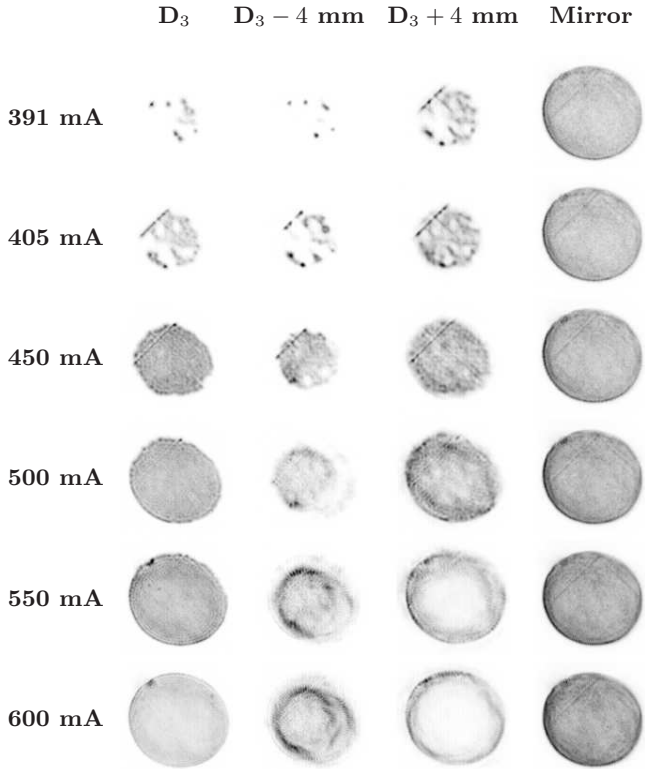}}
\end{picture}
\caption{Near field intensity distribution of the VCSEL with
feedback from a VBG at different currents. Left column: D{\rm $_3$}
at self-imaging distance; center columns: $D_3$ too short or too
large by 4 mm; right column: Feedback with plane mirror.}
\label{fig:Nearfield}
\end{figure}

\begin{figure}[thb]
\includegraphics[width=80mm,clip=]{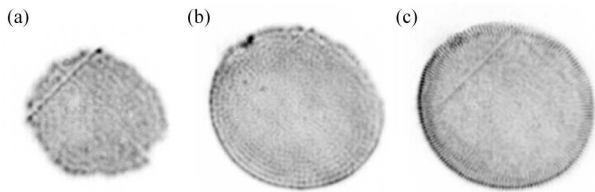}
\caption{Near field intensity distribution of the VCSEL with
feedback from a VBG (a,b) and a plane mirror (c). a, c) $I=450$~mA,
b) $I=550$~mA. } \label{fig:blow-up}
\end{figure}

If the current is increased, the inner part of the lasing aperture
starts to lase also but a certain preference for the perimeter stays
(upper image, right column, Fig.~\ref{fig:Nearfield}).  A blow-up of
a characteristic structure is shown in Fig.~\ref{fig:blow-up}c. If
the current is increased further the output power increases further
but the spatial structure stays essentially the same (lower images,
right column, Fig.~\ref{fig:Nearfield}). The emission in far field (Fig.~\ref{fig:Farfield}, right column)
is quite broad with a disk-shaped structure on axis surrounded by a
faint halo (right column, Fig.\ref{fig:Farfield}). At the transition between center and halo, one wavenumber is somewhat enhanced
leading to a ring (left column, Fig.\ref{fig:Farfield}). This wavenumber is probably favored by the
detuning between the frequency of the gain maximum and the
longitudinal cavity resonance as discussed before for free-running
devices of this kind \cite{schulz-ruhtenberg09a}.

\subsection{Feedback with VBG}

With the VBG the threshold is much higher and at threshold small
localized spots spontaneously appear in the near field of the laser,
away from the boundaries (left column, uppermost image in
Fig.~\ref{fig:Nearfield}). These spots are the laser cavity solitons
(LCS) investigated in \cite{tanguy08,radwell09,ackemann09a}. If the
current is increased further, more LCS appear at other locations and
LCS already formed give way to extended lasing states of lower
amplitude (images in left column of Fig.~\ref{fig:Nearfield}). At
about 500~mA essentially the whole aperture is lasing, whereas with
the mirror this is already the case for about 400~mA. The patterns
are actually quite similar to the ones obtained with a plane mirror,
i.e., fine waves at high spatial frequency filling up the whole
aperture of the device. However, there is the important difference
that the length scale now depends on current: As a comparison
between the blow-ups in Fig.~\ref{fig:blow-up}a and b shows, the
wavelength of these waves decreases with increasing current.

\begin{figure}[thb]
\begin{picture}(85,100)
\put(20,100){\bf {\rm \textbf{D$_3$}}}
\put(30,100){\bf {\rm \textbf{D$_3-4~$mm}}}
\put(50.5,100){\bf {\rm \textbf{D$_3+4~$mm}}}
\put(72,100){\bf {\rm \textbf{Mirror}}}
\put(0,87){\bf 391~mA}
\put(0,71){\bf 405~mA}
\put(0,55){\bf 450~mA}
\put(0,39){\bf 500~mA}
\put(0,23){\bf 550~mA}
\put(0,7){\bf 600~mA}
\put(15,0){\includegraphics[width=70mm,clip=]{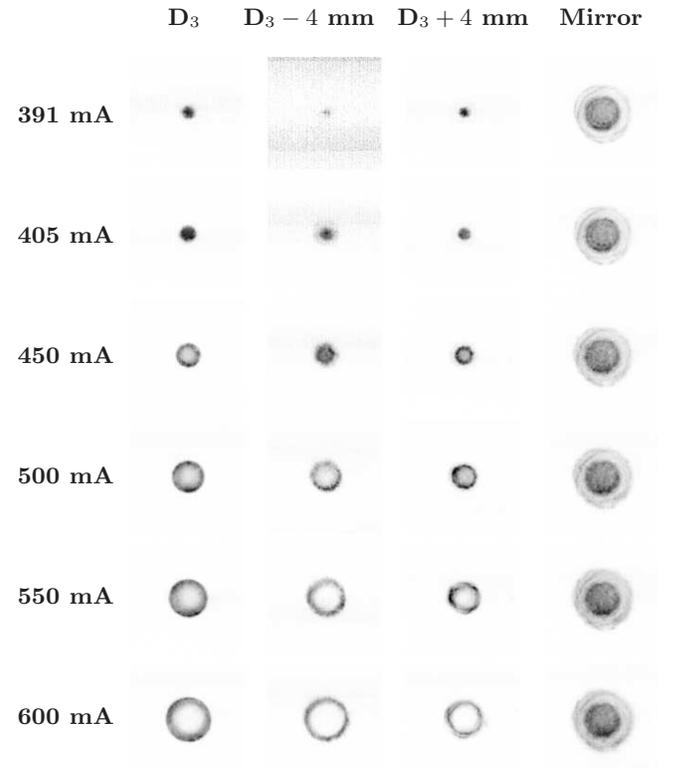}}
\end{picture}
\caption{Far field intensity distribution of the VCSEL with feedback
from a VBG at different currents. Left column: D{\rm $_3$} at
self-imaging distance; center columns: $D_3$ too short or too large
by 4 mm; right column: Feedback with plane mirror.}
\label{fig:Farfield}
\end{figure}

This feature can be much better investigated in far field images.
The sequence depicted in the leftmost column of
Fig.~\ref{fig:Farfield} illustrates that the emission is very
much dominated by a single ring with negligible background, i.e.\
only a single transverse wavenumber is lasing. The solitons start to
emit on axis and the wavenumber increases monotonically with
increasing current. We are going to do a quantitative investigation
at the end of Sect.~\ref{sec:interp}.

If the distance $D_3$ is changed away from the self-imaging
condition the far-field images do not show a significant change
except for quantitative corrections to the wavenumber (analyzed in
more detail in Sect.~\ref{sec:adjust}), but the near-field
images do. For a longer cavity ($+4$~mm) we observe a `defocusing'
effect, i.e., the emission is shifting towards the boundaries of the
aperture, especially at high currents. Inversely, for a too short
cavity  $\rm{-4~mm}$ the pattern seems to be `focused', i.e.\ it
contracts towards the center of the device.

\begin{figure}[thb]
     \centerline{
     \includegraphics[scale=0.45,angle=0]{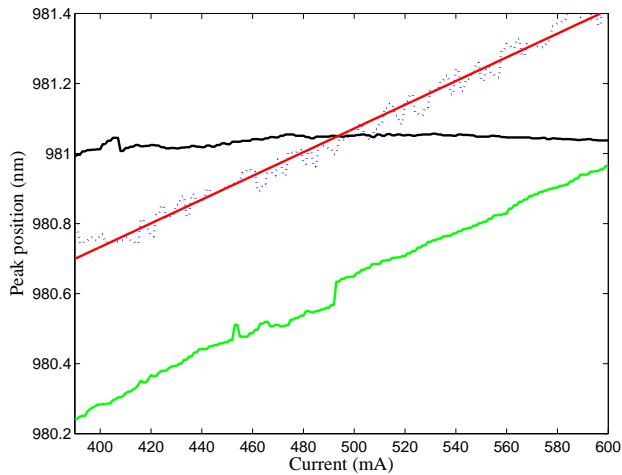}
     }
     \caption{(Color online) Black solid line: Frequency shift of one mode of the VCSEL with feedback of a VBG ($M=6.25$).
     Green solid line (light grey in print): feedback by a plane mirror.
     Blue dashed line(grey in print): Free Running Laser (FRL) at $69~{\rm ^\circ C}$.
     Red solid line (dark grey in print): Fit to the FRL. }
    \label{fig:freq_shift_SC}
\end{figure}

Finally, Fig.~\ref{fig:freq_shift_SC} gives an indication of the
change of emission wavelength with increasing current. The
free-running laser shows an approximately linear increase with a
rate of $0.0035~{\rm nm/mA}$ device. This is due to Joule heating.
The VCSEL with feedback shows a corresponding behavior, whereas the
emission wavelength of the VCSEL with feedback is essentially locked
to one value (within 0.06 nm) given by the peak reflection of
the VBG. This matches qualitatively the observations in EEL
discussed before. A closer inspection shows that the wavelength is
in tendency increasing, by about 0.06~nm, at the beginning (the soliton area) and
then slowly decreasing (by about 0.02~nm).

\section{Interpretation}\label{sec:interp}
\begin{figure}[thb]
     \centerline{
     \includegraphics[scale=0.35,angle=0]{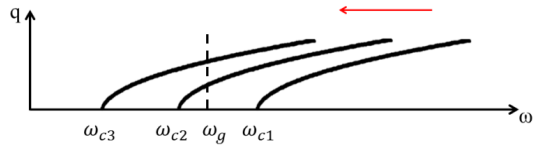}
     }
     \caption{Mechanism for selecting the transverse wavenumber. $\omega_{\rm g}$ : grating frequency. $\omega_{\rm c1}$ : VCSEL frequency at low current. $\omega_{\rm c2}$ : VCSEL frequency at higher current. $\omega_{\rm c3}$ : VCSEL frequency at high current.}
    \label{fig:Redshift}
\end{figure}

The resonance condition for a VCSEL cavity are identical to the ones
of a plano-planar Fabry-Perot cavity in diverging light and were
investigated in detail in
\cite{schulz-ruhtenberg05,schulz-ruhtenberg09a}. The dispersion
relation of plane waves with a transverse wavenumber $q$ in the
VCSEL is given by \cite{schulz-ruhtenberg05,schulz-ruhtenberg09a}:
\begin{equation} \label{q_vcsel}
q_{\rm VCSEL} = \sqrt{\frac{8 \pi^2 n_0 n_{gr}
(\lambda_{\rm c}-\lambda)}{\lambda^3}} = a\sqrt{\Delta \lambda},
\end{equation}

where $ n_0$ is an average refractive index of the VCSEL and
$n_{\rm gr}$ is the group index, $\lambda$ is the vacuum
wavelength of the emission and $\lambda_{\rm g}$  the vacuum
wavelength of the longitudinal resonance. If the wavelength of the
emission is fixed by the VBG, as indicated by
Fig.~\ref{fig:freq_shift_SC}, and the longitudinal resonance shifts
due to the Joule heating, different transverse wavenumbers should
come into resonance with the feedback starting with those at $q=0$.
This situation is schematically depicted in Fig.~\ref{fig:Redshift}. $\omega_{\rm g} = 2\pi c/\lambda_{\rm g}$ represent the grating frequency, $\omega_{\rm c} = 2\pi c/\lambda_{\rm c}$ the longitudinal resonance of the cavity, which decreases with increasing temperature or current.
Due to the dispersion relation of the VCSEL, one expects a selection of not only wavelength, but also of
transverse wavenumber,  i.e.\ the emission should correspond to a
ring in far field, which is exactly what we observed in
Fig.~\ref{fig:Farfield}. The emission angle should increases
monotonically with current. Quantitatively, we expect an increase as
the square root of the detuning of the VCSEL cavity, which we will
investigate below. We mention that close to threshold, around $q=0$,
nonlinear frequency shifts play a role leading to the possibility of
bistability of solitons: Reducing the frequency gap between grating
and VCSEL resonance increases the intensity, which
reduces the carrier density, increases the refractive index and thus
red-shifts the cavity resonance further. This positive feedback can
create an abrupt transition to lasing
\cite{tanguy08,radwell09,naumenko06} and can explain also the
variation in wavelengths in the range below 420~mA in
Fig.~\ref{fig:freq_shift_SC}. If the cavity resonance condition is
not quite homogenous, this takes place at different current levels,
which can explain that the VCSEL starts to lase locally at the
locations where the resonance is most `reddish', i.e.\ closet to the
grating and then slowly fills up as obvious in the near field images
(Fig.~\ref{fig:Nearfield}, left column). We propose to use this
relationship to provide a mapping of the disorder of the cavity
resonance \cite{ackemann12}.

\begin{figure}[thb]
     \centerline{
     \includegraphics[scale=0.40,angle=0]{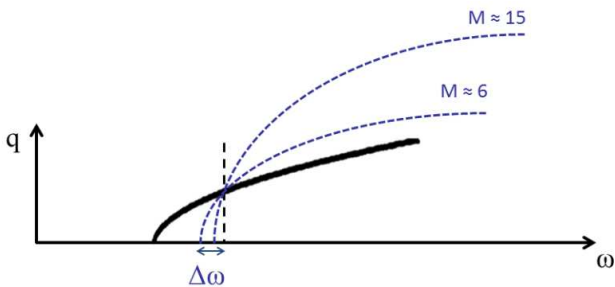}
     }
     \caption{Effect of the magnification $M$ on the dispersion curve of the VBG.
     $\rm{M \approx 6}$ corresponds to the short cavity whereas $\rm{M \approx 15}$ corresponds to the long cavity.
     $\rm{\Delta \omega}$ corresponds to a frequency shift between $\omega_{\rm g}$ and the emission frequency.}
    \label{fig:Magshift}
\end{figure}

The small remaining frequency shift observed at high currents
(larger than 420~mA) in Fig.~\ref{fig:freq_shift_SC} is due to the
fact that the dispersion curve of the VBG is not straight.
That curvature means that there is a small difference between the operating wavelength of the device and the intercept as shown schematically in Fig.~\ref{fig:Magshift}.\\

From the condition that the phase shift in a single layer of the VBG
should remain $\pi/4$ also at oblique incidence, the dispersion
relation of the VBG can be expressed as:

\begin{equation} \label{eq:VBG}
q_{\rm VBG} = M \sqrt{\frac{8 \pi^2 n^2
(\lambda_{\rm g}-\lambda)}{\lambda^3}} = b\sqrt{\Delta \lambda}
\end{equation}

with $\lambda_{\rm g}$ being the peak reflection wavelength of the
VBG and $n$ the refractive index of the glass host. From
Eq.~\ref{eq:VBG}, one can see that if the telescope magnification
$M$ is increased then the dispersion curve will straighten up thus
reducing the frequency shift. By equating (\ref{q_vcsel}) and
(\ref{eq:VBG}), one obtains for $\lambda_{\rm g} \leq \lambda_{\rm c}$ the emission wavelength

\begin{equation}\label{eq:lambda}
\lambda = \frac{\lambda_{\rm g} - \frac{a^2}{b^2} \lambda_{\rm c}}{1 - \frac{a^2}{b^2}} = \lambda_{\rm g} - \frac{\frac{a^2}{b^2} (\lambda_{\rm c} - \lambda_{\rm g})}{1 - \frac{a^2}{b^2}}
\end{equation}

For $M = 6.25$ and for a shift of $\lambda_{\rm c}$ by $\rm 0.0035 nm/mA\times 210 mA = 0.735 nm$, we have $\lambda = \lambda_{\rm g} - {\rm 0.10~nm}$ which is on the order of -- though somewhat larger -- what we observe in Fig.~\ref{fig:freq_shift_SC}.

\begin{figure}[thb]
     \centerline{
     \includegraphics[scale=0.45,angle=0]{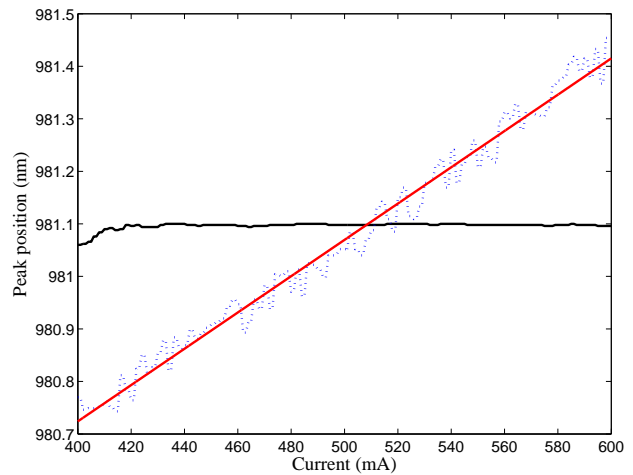}
     }
     \caption{(Color online) Black solid line: Frequency shift of the VCSEL with feedback of a VBG ($M=15.6$). Blue dashed line (light grey in print): FRL at $16~{\rm ^\circ C}$. Red solid line (grey in print): Fit to the FRL. }
    \label{fig:Freq_shift_LC}
\end{figure}

To test the prediction of Eq.~\ref{eq:lambda}, the same experiment
has been done with a second telescope arrangement i.e.\, a long
cavity with $f_2=125$~mm ($M \approx 15$) and
Fig.~\ref{fig:Freq_shift_LC} illustrates the shift in frequency of
the VCSEL with this external cavity setup. The shift is barely
noticeable and within the accuracy range of our equipment it is not
possible to quantify it. From Eq.~\ref{eq:lambda} one expects to
have $\lambda = \lambda_{\rm g} - {\rm 0.02~nm}$ with $M = 15.6$. This
result is in good agreement with our experimental observation that the curve is essentially flat beyond the soliton region (up to about 430~mA). The remaining variations are below the limit of the accuracy of our OSA (about 0.03~nm relative accuracy within a single run). These results confirm the
presence of frequency locking when feedback is provided from a VBG
and also that the remaining frequency shift depends on the
magnification factor of the external cavity. For large magnification ($M >>1$), $b >>a$, one finds $\lambda \approx \lambda_{\rm g}$, i.e.\ perfect locking because the dispersion relation of the VBG approaches a vertical line.

\begin{figure*}[bt]
\begin{picture}(170,68)
\put(0,68){\bf a)} \put(85,68){\bf b)}
\put(0,0){\includegraphics[width=82mm,clip=]{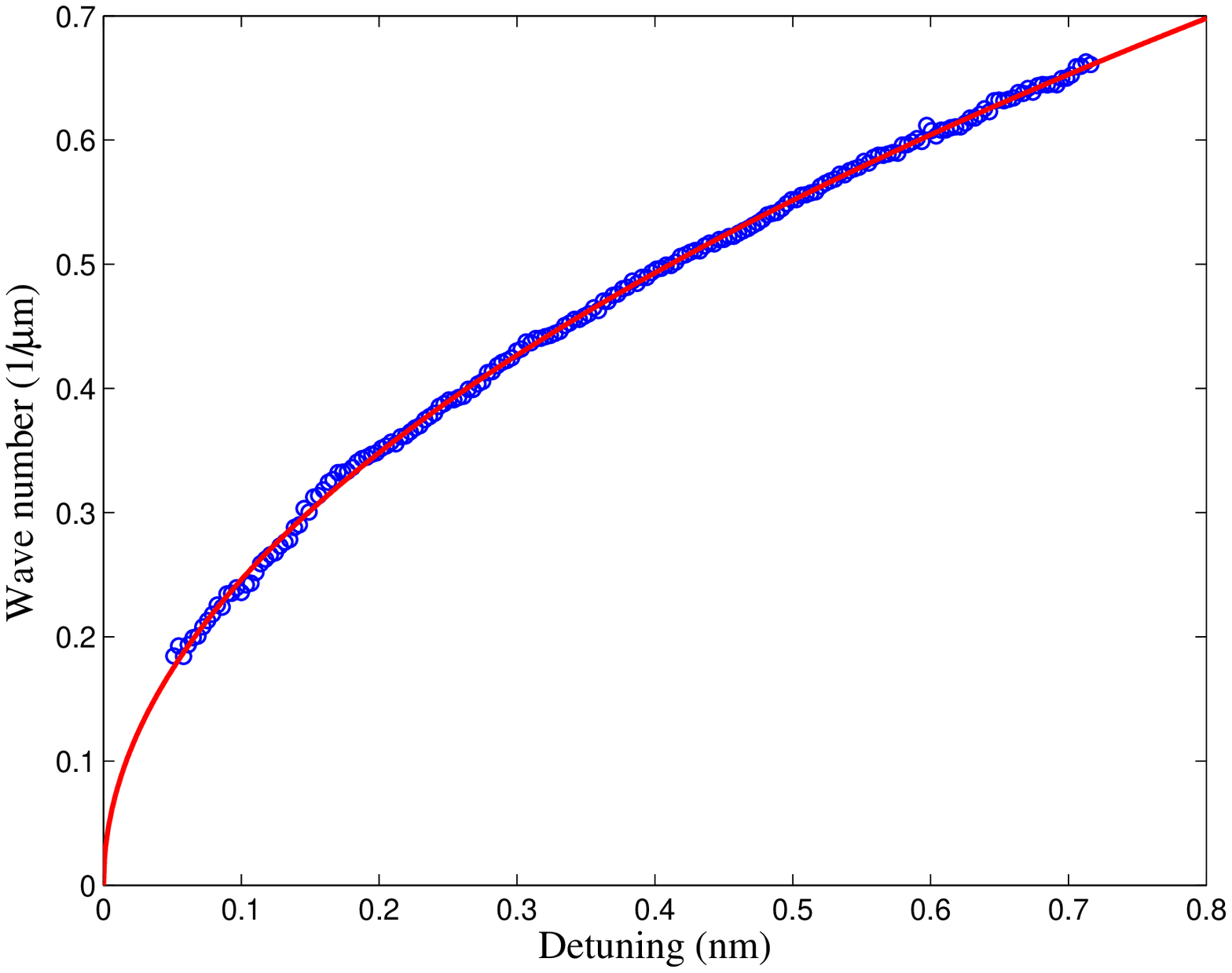}}
\put(85,0){\includegraphics[width=82mm,clip=]{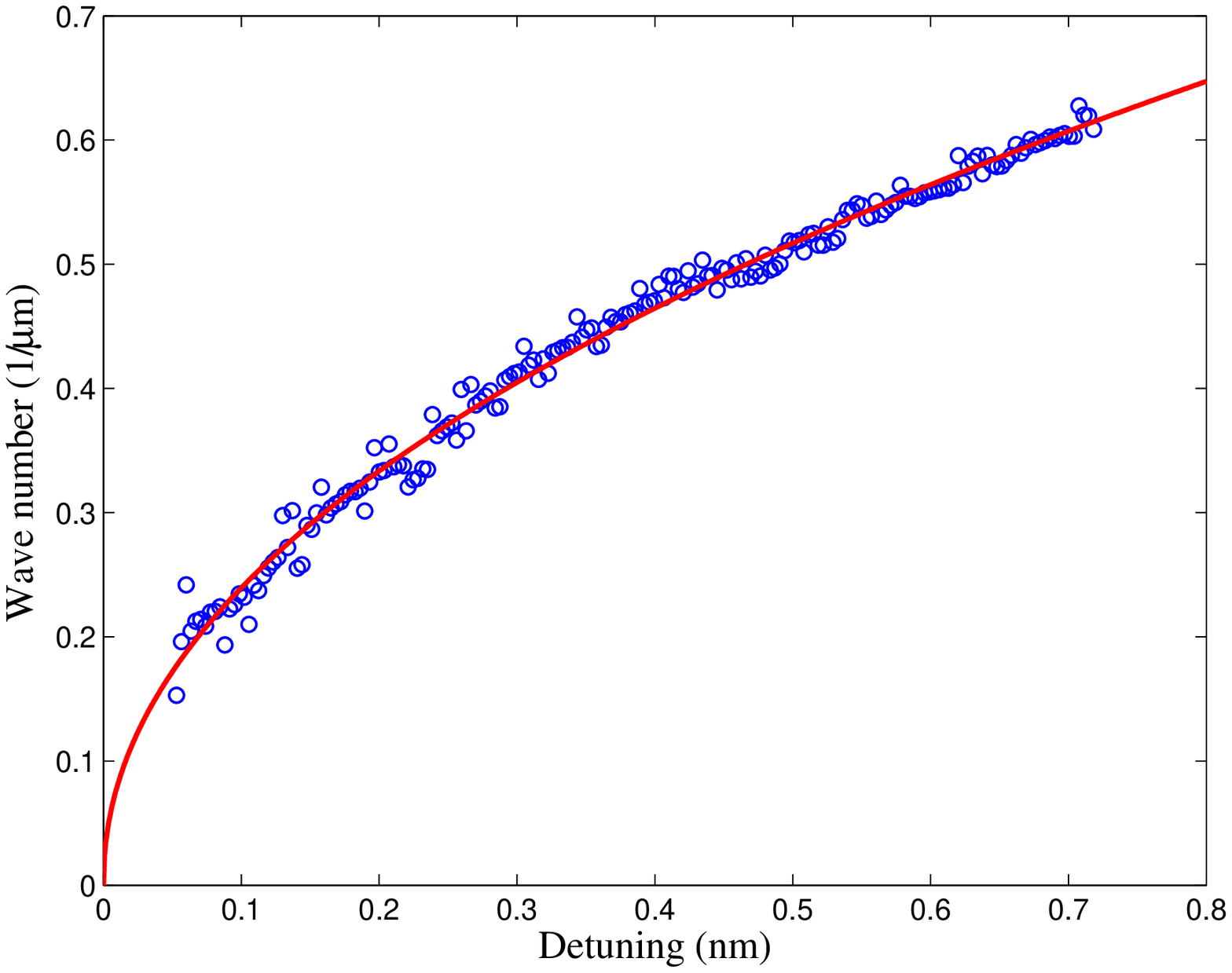}}
\end{picture}
\caption{Transverse wave number of the emission versus detuning between emission wavelength and longitudinal resonance (crosses represent data points and solid lines fits).
a) For $\rm{D_2}$=70.5~mm leading to a scaling exponent $s=0.502$; b)
for $\rm{D_2}$=78 mm leading to a scaling exponent $s=0.479$.
These plots are produced by finding the intersect $I_0$ and the scaling exponent $s$ using a fitting function
$q(I)=c(I-I_0)^s$. The intercept $I_0$ then corresponds to the zero
detuning condition. Starting from that the rate of
$\rm{0.0035~nm/mA}$ is applied as a coefficient to define the
detuning scaling, which is used to plot the fitting curve (red curve
in the figure above). The programm also returns the $\rm{R^2 value}$
for the fit.} \label{fig:Transverse}
\end{figure*}

As derived in the previous section, the wavenumber of the emission
should follow a square root behavior with detuning. Based on data
like the ones obtained in Fig.~\ref{fig:Farfield} this can be
checked quantitatively and the results are shown in
Fig.~\ref{fig:Transverse}a, which shows indeed a nice square-root
behavior with a scaling exponent of $0.502$ with a negligible
parameter error (on the order of $3\e{-3}$) and a high-fidelity regression coefficient ($R^2=0.9989$) returned by the nonlinear fitting code.

\section{Effects of deviation from self-imaging}\label{sec:adjust}

By these investigations, the effect of the VBG on wavelength locking
is quite clear but the influence of the distances between the
intra-cavity lenses (e.g.\ $\rm{D_2}$ or $\rm{D_3}$) remains to be
investigated. For example, Figs.~\ref{fig:Nearfield} and
\ref{fig:Farfield} clearly indicate that $D_3$ has an influence on
the spatial structures.

As shown in Fig.~\ref{fig:setup}, a self imaging cavity is
characterized by three distances. The first one is the distance
$\rm{D_1}$ between the laser and the collimation lens. It is found
by adjusting the position of the lens until the divergence of the
beam is minimal. The second distance $\rm{D_2}$ is the distance
separating both lenses also referred as the intra-cavity distance in
the following; ideally it is $(f_1 + f_2)$. The last distance is the
distance $\rm{D_3}$ separating the second lens and the VBG. It is
determined by moving the VBG until the near field image of the VCSEL
has the sharpest boundaries. As indicated, matrix theory indicates
that $D_2$ is not influencing the imaging condition for the
intensity distribution. Since $D_2$ is hence the most poorly defined
one (in the sense that there is no obvious alignment criterion like for $\rm{D_1}$, $\rm{D_3}$) we first turn our attention to it.

Fig.~\ref{fig:Transverse}b shows a dispersion relation similar to the one in
a) but obtained at $D_2=78$~mm, away from the self-imaging
condition. Though the data still seem to follow roughly a square-law like behavior,
they are more scattered and the lower quality of the fit is evidenced also by reduction of the regression coefficient to $\rm{R^2=0.9845}$ and the parameter error returned is $11\e{-3}$. The scaling exponent turns out to be slightly different from 0.5, e.g.\ 0.479 in Fig.~\ref{fig:Transverse}b. This scattered behavior combined with a reduction of the fitting quality and a deviation of the scaling exponent from 0.5 is typical also for other distances $\rm{D_2}$ different from $D_2=70.5$~mm. For example, for $D_2=74$~mm, the scaling exponent is 0.508 and $\rm{R^2=0.9874}$.

\begin{figure}[hbt]
     \centerline{
     \includegraphics[scale=0.45,angle=0]{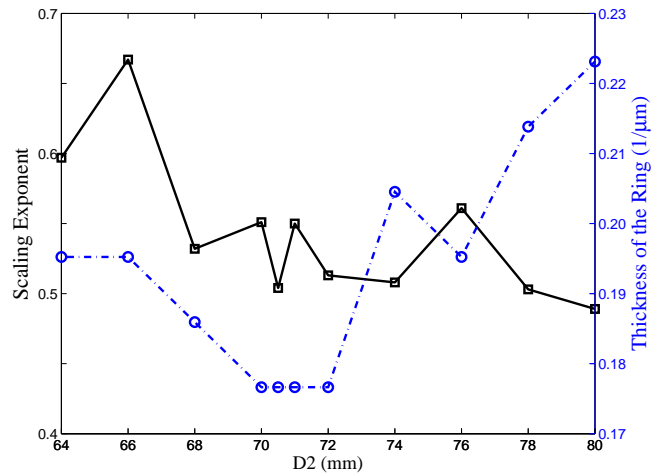}
     }
     \caption{Black solid line: Scaling exponent vs.\ $D_2$. Blue dashed line: Thickness of the ring in far field  vs.\ $D_2$.}
    \label{fig:qD2}
\end{figure}

We also notice that there is an enhanced tendency for slight variations of the details of the patterns (not the basic structure) from run to run away from $D_2=70.5$~mm. Hence we report results averaged over three runs in Tab.~\ref{table:scaling_exp}, which summarizes the effect of the intra
cavity distance on the scaling exponent over a larger range. Fig.~\ref{fig:qD2} illustrates the same results graphically.
The first observation is that in general the averaged scaling exponent is
oscillating around 0.5 with a tendency to larger values for smaller distances, but that there are actually two positions at
$D_2=74~\rm{and}~78$~mm in addition to $D_2=70.5$~mm at which the averaged scaling exponent is again close to 0.500. However, as stated already above, the quality of the fits is lower and there is a larger variation between runs which we characterize by the standard deviation $\sigma$ of the scaling exponents returned for the different runs (Tab.~\ref{table:scaling_exp}, last row).

In order to get further insight in the nature of these deviations and variations, we analyzed the width of the ring in Fourier space as a measure of the quality of wavenumber selection (Tab.~\ref{table:scaling_exp}, 2nd row and Fig.~\ref{fig:qD2}, dashed line). It has a minimum around $D_2=70.5$~mm, i.e.\ the anticipated self-imaging condition,  and increases, except for one exception interpreted as scatter, monotonically away from it. In particular, at 70.5~mm the ring is narrower than at both 74 mm and 78 mm (see also the width in Tab.~\ref{table:scaling_exp}). These observations reinforce our expectation that $D_2=70.5$~mm is the self-imaging position.

\begin{table*}
\caption{Effect of $\rm{D_2}$ on the scaling exponent. Three rounds
of data taking averaged. The data for both 70.5 and 72 are actually based on nine runs. The width of the ring in far field is given
at 500 mA.} \centering \setlength{\tabcolsep}{7pt}
\begin{tabular}{c c c c c c c c c c c c}
\hline\hline
Distance $\rm{D_2}$ (mm) & 64 & 66 & 68 & 70 & 70.5 & 71 & 72 & 74 & 76 & 78 & 80 \\ [0.5ex]
\hline
Scaling Exponent & 0.597 & 0.667 & 0.532 & 0.551 & 0.504 & 0.550 & 0.513 & 0.508 & 0.561 & 0.503 & 0.489 \\
Width (1/${\rm\mu m}$) & 0.195 & 0.195 & 0.186 & 0.176 & 0.176 & 0.176 & 0.176 & 0.204 & 0.195 & 0.214 & 0.223 \\
Intersect (mA) & 389 & 381 & 395 & 392 & 395 & 393 & 396 & 391 & 384 & 394 & 397 \\
St. deviation $\sigma$ & 0.022 & 0.011 & 0.013 & 0.011 & 0.002 & 0.032 & 0.022 & 0.010 & 0.022 & 0.043 & 0.009 \\ [1ex]
\hline
\end{tabular}
\label{table:scaling_exp}
\end{table*}

This behavior can be understood by the ABCD-matrix for the
round-trip through the external cavity if $D_2=f_1+f_2+x$, which
reads
\begin{equation}\label{eq:D2}
M_2 = \left(\begin{array}{cc} 1 & 0 \\ - \frac{2}{f_1^2}\,x &
1\end{array}\right) . \end{equation} A small mistuning $x$ leads to
a change of phase curvature or ray angle for the returning light.
Hence the spatial Fourier spectrum broadens. Hence we conclude that
$\rm{D_2}$=70.5~mm corresponds to the self-imaging distance. It is
also within $\pm 0.2$~mm of the best estimate we can
give on the expected cavity length from the dispersion data and the
knowledge of the principal planes available for the lenses used.

In order to confirm that, another experiment has been set up to
study the cavity in details. It consists of the same telescope with
the exact same optics (a collimation lens, a focusing lens and a
beam splitter). A tunable laser is coupled into a single mode fibre.
The collimated output beam is then injected into the telescope from
the side with the larger focal length lens. One should expect that
the beam coming out of the telescope at the collimation lens has the
smallest beam divergence if the intra-cavity distance is right and
indeed best collimation is found within $\pm 0.2$~mm
of $\rm{D_2}$=70.5~mm, which is considered to be the accuracy of the
distance measurement.

\begin{figure}[htb]
     \centerline{
     \includegraphics[scale=0.45,angle=0]{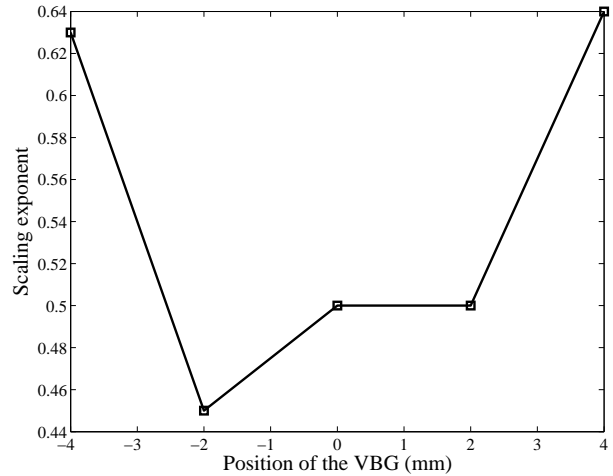}
     }
     \caption{Scaling exponent as function of $\rm{D_3}$. The 0 position indicates self imaging position and negative sign means that the VBG is moved towards the telescope. Inversely, the positive sign means that the VBG is moved away from the focusing lens.}
    \label{fig:PosVBG}
\end{figure}

Finally, we discuss the influence of $\rm{D_3}$ on the pattern formation.
Fig.~\ref{fig:PosVBG} indicates that it also has an influence on the
scaling exponent as well as on the manifestation of structures. We noticed also in Fig.~\ref{fig:Nearfield} that
for a too long cavity the near field seems to be pushed to the
perimeter, whereas for a short cavity it seems to contract inwards.
The boundaries are not as sharp and well defined. Again these
features can be understood by ABCD-matrix theory, which gives for
$D_3=f_2+x$ a round trip matrix of
\begin{equation}\label{eq:D3}
M_2 = \left(\begin{array}{cc} 1 & \frac{2f_1^2}{f_2^2}\ x \\ 0 & 1\end{array}\right) ,
\end{equation}
i.e.\ the system is not imaging any more. For a ray emitted at a
radius $r$ at an angle $\Theta$, the returning rays hits at
\begin{equation} r' = r + \frac{2f_1^2}{f_2^2}\ \Theta x .\end{equation}
This means that for $x>0$ (too long cavity), the emission is pushed
outwards, whereas for $x<0$ (too short cavity) it is pushed inward.
This is the tendency observed in the experiment. Note that the angle
$\Theta$ does not change very much.

\section{Conclusion}
We established that VBGs work as wavelength lockers for VCSELs as well as for EEL lasers for which they are of significant importance for wavelength and modal control. VCSELs have already a lower temperature dependence than
 EEL because their wavelength shift is determined by the cavity shift of about 0.1~nm/K (GaAs based devices) compared to the gain shift of 0.2-0.3~nm/K relevant to EEL. Nevertheless, an external VBG can decrease this shift even more and the fidelity increases with increasing magnification of the imaging system, unfortunately increasing the foot print of the system.

 For increasing temperature or current, the on-axis modes come into resonance first and here phase-amplitude coupling leads to the possibility of bistability \cite{naumenko06} and, via self-focusing, soliton formation \cite{tanguy08,radwell09,ackemann09a}. Beyond threshold, we can stabilize single-wavenumber and narrow bandwidth emission with fairly high fidelity. These structures might find applications where ring-shaped foci are desired. The radius of the ring can be tuned by current (or temperature). This feature constitutes an important difference between VCSELs and EEL and stems from the fact that VCSELs are intrinsically single-longitudinal mode devices. Hence the wavelength locking is accompanied by an increase in mode order (transverse wavenumber) for increasing temperature (induced either directly via the ambient temperature or via Joule heating). In contrast, for an EEL consequent longitudinal orders will shift into resonance and hence the modal distribution will contain the mixture of low-order and high-order modes typical for broad-area EEL, but the average mode order won't be increasing so much as in the VCSEL.

 We also analyzed the effect of deviations of the external cavity from the self-imaging condition. It turns out that the system is remarkably insensitive to deviations even in the mm range for a cavity length in the 100~mm range. This is good news for the robustness of experimental conditions. We remark that we consider the self-imaging condition to be attractive for these kind of experiments because it allows for best feedback efficiency in terms of amplitude and phase. An autocollimation setup using only the collimation lens (as in \cite{marino03,chembo09}) not only introduces an image inversion (relevant for Gauss-Hermite modes for example), but due to the shortness of the focal lengths of typical collimation lenses, it is also likely that the length of the external cavity is larger than the perfect telescope length (an imperfect $D_2$ in our terminology). Hence, there is a phase curvature of the returning beam for Gaussian modes and a chance of ray direction for Fourier modes and as a result an imperfect interference with the cavity mode.  Deviations from the self-imaging conditions make it also more difficult to model the returning field distribution, though in principle methods like the Collins-integral are available, of course \cite{dunlop98}.


\section*{Acknowledgements}
Y.N. is supported by an EPSRC DTG. We are grateful to W.J. Firth and Jesus Jimenez-Garcia for useful discussions and to R. Jaeger (Ulm Photonics) for supplying the devices.



       %

\end{document}